\documentclass[fleqn,usenatbib]{mnras}

\usepackage{newtxtext,newtxmath}
\usepackage{xcolor}
\usepackage[normalem]{ulem}
\usepackage{hyperref}

\usepackage[T1]{fontenc}

\DeclareRobustCommand{\VAN}[3]{#2}
\let\VANthebibliography\thebibliography
\def\thebibliography{\DeclareRobustCommand{\VAN}[3]{##3}\VANthebibliography}

\def\nu{NuSTAR}

\usepackage{graphicx}	% Including figure files
\usepackage{amsmath}	% Advanced maths commands
\title[Wavelet-based method for NuSTAR stray light studies]{Wavelet-based image decomposition method  for  NuSTAR stray light background studies}

\author[Mukhin et al.]{
Andrey Mukhin,$^{1,2}\thanks{Corr. author. E-mail: amukhin@cosmos.ru}$ Roman Krivonos,$^{1,2}$ Alexey Vikhlinin,$^{3,1}$
Brian Grefenstette,$^{4}$,
\newauthor Kristin Madsen,$^{5}$ and Daniel Wik$^{6}$\\
$^{1}$Space Research Institute (IKI), Russian Academy of Sciences, Moscow 117997, Russia\\
$^{2}$Institute for Nuclear Research, Russian Academy of Sciences, Moscow 117312, Russia\\
$^{3}$Harvard-Smithsonian Center for Astrophysics, 60 Garden Street, Cambridge, MA 02138, USA\\
$^{4}$Space Radiation Laboratory, Caltech, 1200 E California Blvd, Pasadena, CA 91125, USA;\\
$^{5}$CRESST and X-ray Astrophysics Laboratory, NASA Goddard Space Flight Center, Greenbelt, MD 20771, USA\\
$^{6}$Department of Physics and Astronomy, University of Utah, Salt Lake City, UT 84112, USA;\\
}

\date{Accepted in JATIS SPIE 10 October 2023. Published: 26 October 2023. DOI: \url{https://doi.org/10.1117/1.JATIS.9.4.048001}}

\pubyear{2023}

\begin{document}
\label{firstpage}
\pagerange{\pageref{firstpage}--\pageref{lastpage}}
\maketitle

\begin{abstract}
The large side aperture of the {\nu} telescope for unfocused photons (so-called stray light) is a known source of rich astrophysical information. To support many studies based on the {\nu} stray light data, we present a fully automatic method for determining detector area suitable for background analysis and free from any kind of focused X-ray flux. The method's main idea is `\'a trous' wavelet image decomposition, capable of detecting structures of any spatial scale and shape, which makes the method of general use. Applied to the {\nu} data, the method provides a detector image region with the highest possible statistical quality, suitable for the {\nu} stray light studies. We developed an open-source Python \texttt{nuwavdet} package, which implements the presented method. The package contains subroutines to generate detector image region for further stray light analysis and/or to produce a list of detector bad-flagged pixels for processing in the {\nu} Data Analysis Software for conventional X-ray analysis.
\end{abstract}

% Include a list of up to six keywords after the abstract

%%%%%%%%%%%%%%%%%%%%%%%%%%%%%%%%%%%%%%%%%%%%%%%%%%

%%%%%%%%%%%%%%%%% BODY OF PAPER %%%%%%%%%%%%%%%%%%

\section{Introduction}
\label{intro}
The Nuclear Spectroscopic Telescope Array (\nu) \citep{2013Harrison} is a focusing hard X-ray orbital telescope designed to study point-like and moderately extended (with a few arc minutes in size) X-ray emitters. Due to the coating material of the Wolter Type I grazing-incidence optics, {\nu}'s working energy band is limited by $78.4$~keV, which is the location of the mirrors' Platinum K absorption edge. Most of the astrophysical studies have been performed based on the photons collected within the $13'\times13'$ focused field of view (FOV) of the telescope. However, a very low internal detector background \citep{2013Harrison} is contaminated by photons bypassing the optics and falling directly on the detector array (so-called stray light)\citep{2014Wik, 2017Madsen}. As shown below, this unfocused background can act as a source of rich astrophysical information.

The most direct way to use the unfocused background is to investigate photons from a bright source that is located outside the FOV ($1-4^{\circ}$)\citep{2017Madsen}. This approach allows the extension of the working energy range beyond the mirror's absorption edge \citep{2022ApJ...941...35M} and provides a simplistic instrument response thanks to a good absolute calibration of the detectors, which allowed to measure the absolute intrinsic flux of the Crab at an accuracy better than $4\%$\citep{2017ApJ...841...56M}. The authors of that work also obtained direct measurements of the detector absorption parameters without the added complications of the mirror response. Other examples of bright stray light sources are presented in the on-going catalog StrayCats \citep{2021ApJ...909...30G, 2022Ludlam}, the catalog of the {\nu} observations containing the unfocused flux from the bright sources. It was used to perform the first scientific timing analysis of a stray light source \citep{2022ApJ...926..187B}.

Another interesting source of information is the flux from the population of weak X-ray sources outside the FOV of the telescope that form an additional background (which we refer to as stray light background) at the detector plane of the telescope and usually are not resolved to individual sources. The {\nu} stray light background contains information about the surface brightness of the X-ray sky up to ${\sim}4^{\circ}$ from the optical axis. This property has been used \citep{2019Perez} to study the diffuse hard X-ray emission in the Bulge of the Milky Way, broadly known as Galactic Ridge X-ray Emission (GRXE \citep{2006AandA...452..169R, 2007AandA...463..957K, 2012ApJ...753..129Y}) and to obtain the first measurement of the Cosmic X-ray Background (CXB) in $3-20$~keV energy band using the {\nu} stray light background \citep{2021Krivonos}. In this article the authors successfully separate stray light and detector instrumental background, thus obtaining the X-ray spectrum of the CXB based on the information of a broad area of the sky (see also \citep{2023AJ....166...20R}).

Aside from the analysis of CXB properties, the flux of stray light background can be used to search for X-ray lines from the radiative decay of sterile neutrino Dark Matter. {\nu} set strong limits on the sterile neutrino mass and mixing angle finding no line detection in the blank-sky extragalactic fields \citep{2016PhRvD..94l3504N}; the Galactic Center \citep{2017PhRvD..95l3002P}, the Bulge \citep{2020Roach} and the halo \citep{2023PhRvD.107b3009R}; and the direction of M31 galaxy \citep{2019PhRvD..99h3005N}.

The main idea of taking advantage of {\nu} stray light background is to characterize specific detector spatial variations, caused by side-aperture flux. To do this, one needs to clean the detector image of any focused light, usually cast by X-ray sources. 

In an ideal case, the flux contribution from a point-like X-ray source can be modeled with the known Point Spread Function (PSF) of the telescope optics, and the exclusion region can be simply determined by drawing some large fraction of it. However, detailed PSF {\it parametric} model fitting, including distorted off-axis PSF shapes, for each observation containing a point source requires considerable effort and can not be implemented fully automated. The latter becomes critically important since a large number of observations are required to analyze stray light background. Furthermore, blended or extended X-ray sources, {\nu}'s mast movement resulting in the shifts of the focal point, ghost-rays (photons that only undergo a single reflection inside the optics), and other observational artifacts \citep{2017Madsen} make detailed PSF modeling seriously complicated or practically impossible.

Instead of implementing parametric PSF fitting to clean {\nu} detector from any focused light with any, even irregular shapes (including the ghost-rays contamination), we suggest utilizing the non-parametric method based on wavelet image decomposition. The method is fast, fully automatic, well calibrated, and hence, is suitable for analyzing large data sets.

In this article we present an algorithm for determining the region on the {\nu} detector free from any focused light, and suitable for analyzing {\nu} stray light background. The method can also be used in conventional data analysis to define the source-free region for background extraction. To do this, we provide open Python-based code that creates bad-pixel matrices compatible with the {\nu} data analysis software.

\section{{\nu} stray light background}
\label{sec:nubkg}
The {\nu} observatory carries two co-aligned X-ray telescopes, with an optics bench housing nested grazing type mirror shells, that focuses X-ray photons toward corresponding Focal Plane Modules A and B (FPMA and FPMB). The optics bench is mechanically connected to the FPMs by an open mast, which provides a 10-meter focal plane distance for the entire optical scheme of the telescope. Due to this open geometry, FPMs can register unfocused light from far off-axis angles $1-4^{\circ}$  \citep{2017Madsen}. This side aperture flux from different azimuth directions around the optical axis puts the {\nu} into collimator mode and introduces so-called stray light background \citep{2014Wik}.

As the stray light directly illuminates the {\nu} detectors, the native coordinate reference system for its study is a physical detector plane rather than a sky reference system used for focused light. This coordinate system is represented by a $64 \times 64$ physical detector pixel grid (referred to as RAW), one for each FPM. RAW coordinate image is then transformed into the grid of $360\times360$ pixels (referred to as DET1), which provides better spatial resolution by changing pixel size by a factor of $1/5$, thanks to calibrated probability distribution functions of each physical pixel, based on event grades (see the {\nu} software guide \url{https://heasarc.gsfc.nasa.gov/docs/nustar/analysis/nustar_swguide.pdf}, Sec.~3.9.1 for details). DET1 images are transformed from RAW with the {\nu} data analysis tool \texttt{nucoord}, a part of {\nu} Data Analysis Software (NuSTARDAS). In this work we use DET1 coordinate system unless otherwise stated.

{\nu} stray light background forms a certain spatial pattern on the detector plane, which is a result of the convolution of the sky and angular response function of {\nu} as a collimated instrument modified by the optical bench structure \citep{2014Wik}. The generated pattern is different for the two telescope modules since they have different views on the optical bench.

\begin{figure}
    \centering
    \includegraphics[width=\linewidth]{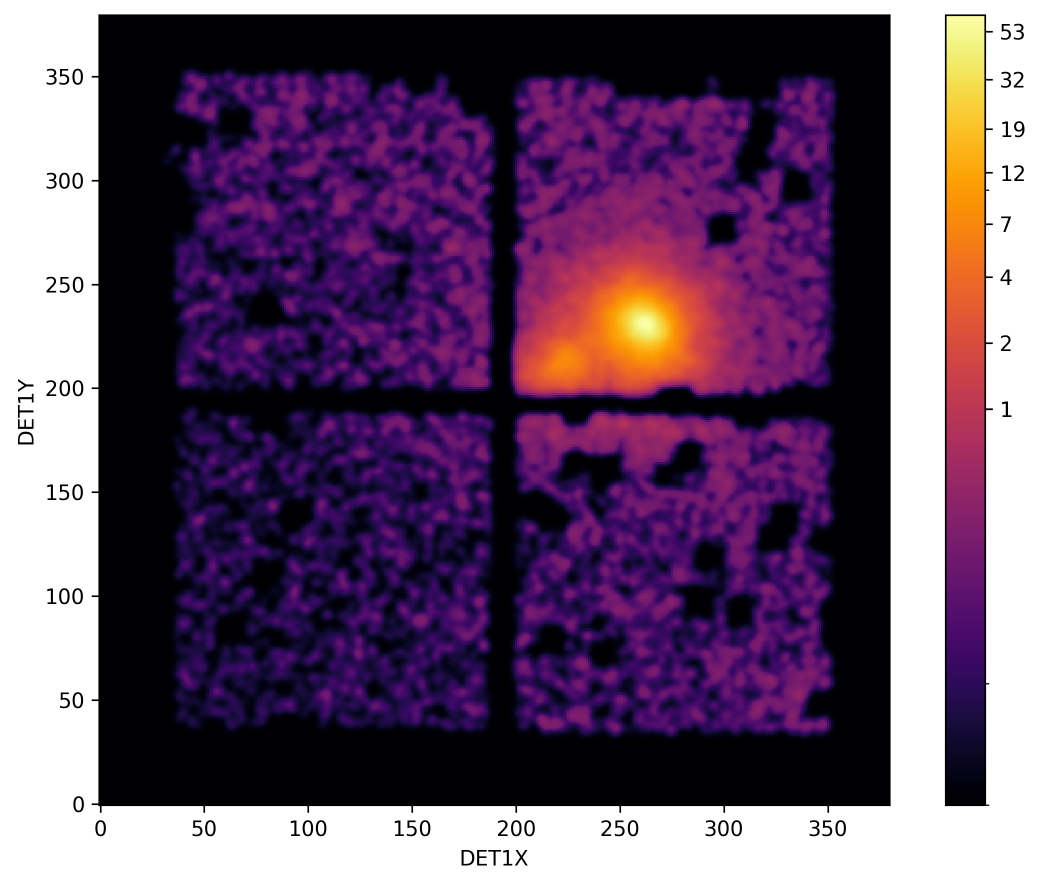}
    \caption{The {\nu} 3$-$20~KeV image of a bright X-ray source in DET1 coordinate system (ObsID 30001019002, FPMB). Hereafter, the images were smoothed with Gaussian kernel $\sigma=2$ pix for a better visual view. This image is used for reference in the following.}
    \label{fig:observation}
\end{figure}

Figure \ref{fig:observation} shows a typical {\nu} observation in DET1 coordinate system with a bright source in the FOV. The X-ray source is a known intermediate polar V2731~Oph \citep{2019ApJ...880..128L,2019MNRAS.482.3622S} observed with {\nu} for ${\sim}50$~ks on 2014 May 14 (ObsID 30001019002, only FMPB is shown). This observation has two notable features. First, despite the only one source was observed on-axis, the DET1 image shows a secondary maximum in addition to the main peak. The reason is a shift of the optical axis due to re-pointing of the telescope while switching between onboard star-tracking systems. Second, the observable spatial background gradient formed by the stray light flux, which we are trying to preserve for further study. We will use this observation as a reference image for the demonstration of the method presented in this paper. The further description does not involve the discussion of different energy ranges, but the choice of energy range can be modified in the analysis depending on it's goals. In this work and the further demonstrations we restricted the energy range to 3 - 20 keV. The choice of this energy range is supported by the fact that the maximum of effective area of {\nu} is located within this range, while at energies above 20 keV the emission lines due to instrumental fluorescence and
activation from SAA passages dominate the background. Additionally, 3 - 20 keV is a working energy range for the CXB studies with the {\nu} stray light data\citep{2021Krivonos, 2023AJ....166...20R}, which we assume to be one of the main applications for the provided algorithm.

\section{Method description}
\subsection{Wavelet decomposition}
\label{sec:method}

In this section we present a method developed for detection and removal from the {\nu} detector image of any type of focused signal characterized by different shapes and angular sizes.

Prior to the actual application of the method to the {\nu} observation, we first generate the images of the observation in DET1 coordinate system and run image correcting procedures to handle specific {\nu} instrumental image artifacts, including dead time correction for bad pixels, detector gaps and different relative chip normalization. A detailed description of the procedure is presented in Appendix \ref{app:prep}. As a result the original data is complemented with the locally simulated count rate filling in the gaps and dead pixels to produce the continuous image preserving the original structures. 

To detect spatial structures on the image we use the so-called \'a trous digital wavelet transform algorithm \citep{1994A&A...288..342S,1994AJ....108.1996S, 1997ApJ...474L...7V, 1998ApJ...502..558V}, which was successfully applied, for example, to {\nu} and INTEGRAL observations \citep{2010A&A...519A.107K, 2014ApJ...781..107K}. Our implementation of the method is based on \texttt{wvdecomp} public code\citep{avikhlinin_2020_3610345, 2020ApJ...892...34S}.

\begin{figure*}
    \centering
    \includegraphics[width=\linewidth]{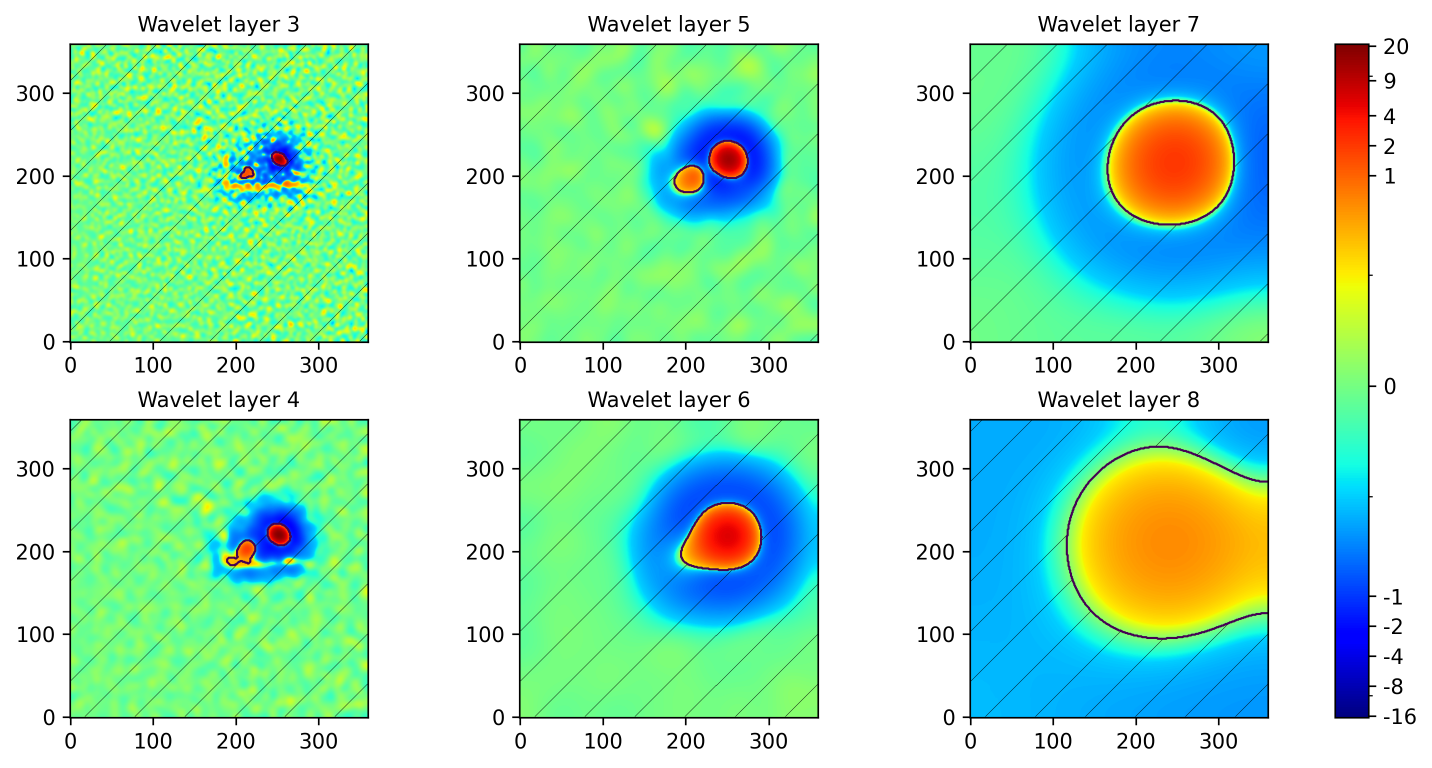}
    \caption{Wavelet image decomposition of a reference {\nu} observation with a bright source in the FOV (see Fig.~\ref{fig:observation}). Images represent wavelet layers at a given scale from 3 to 8. Black contours show significantly detected structures with conditional threshold filtering. The hatched area represents the region with no significantly detected spatial structures. ($t_{\rm max}=3$ and $t_{\rm min}=2$, see Appendix ~\ref{app:threshold})}
    \label{fig:wav_layers}
\end{figure*}

The main advantage of \'a trous algorithm over other wavelet methods is the fact that it decomposes the image into a linear combination of layers or “scales” $w_i(x, y)$. Each of the layers emphasizes the structures with a characteristic size of $2^i$. The overview of the general approach for linear decomposition of 2D data using wavelet transform is described in Appendix \ref{app:wavelet}.

Table~\ref{tab:sizes} shows the characteristic size of wavelet scales $1-8$ expressed in physical and angular units calculated for the size of the {\nu} detector pixel, which is 0.6 mm ($12.3''$) in RAW coordinate system, correspondingly scaled by a factor of 5 for DET1.

The pixel spatial non-uniformities described in RAW to DET1 conversion procedure (Section ~\ref{sec:nubkg}) have size of 5 DET1 pixels and are characterized by wavelet scales 1-2. The structures of such scales will not interfere with wavelet decomposition at scale i=3 (and above), which corresponds to 8 DET1 pixels.

Given the $18''$ angular resolution (full width at half maximum, FWHM) of the {\nu} Point Spread Function (PSF) and its large wings characterized by a half-power diameter (HPD), enclosing half of the focused X-rays $60''$ \citep{2015ApJS..220....8M}, one can expect that point-like X-ray sources will contribute at wavelet scales $3-5$. However, for the observations with bright sources we noticed significant contribution up to scale 8 as shown by wavelet decomposition of a reference image in Fig.~\ref{fig:wav_layers}. For this reason, we extended range of wavelet scales to 3-8 as containing the focused X-ray component.

The characteristic spatial scale of the stray light background, typically cast by CXB on the detector, is comparable with the full size of the detector plane (360 DET1 pixels, ${\sim}40$~mm or ${\sim}13'$), which allows it not to interfere with focused component and remain practically unaffected through the wavelet decomposition at scales 3-8.
\begin{table}
    \begin{center} 
        \caption{Spatial scale of wavelet planes expressed in the {\nu} DET1 pixel (`pix'), physical (`mm'), and angular size ($''$).}
        \label{tab:sizes}
        \begin{tabular}[width=\textwidth]{|l|r|r|r|r|r|r|r|r|r|r}
        \hline
            Scale & \multicolumn{3}{|c|}{Size} \\ 
             & (pix) & (mm) & ($''$) \\ 
        \hline
            1 & 2  & 0.24 & 4.92 \\
            2 & 4  & 0.48 & 9.84 \\
            3 & 8  & 0.96  & 19.68 \\
            4 & 16  & 1.92  & 39.36\\
            5 & 32  & 3.84 & 78.72\\
            6 & 64  & 7.68 & 157.44 \\
            7 & 128  & 15.36 & 314.88 \\
            8 & 256  & 30.72 &   629.76  \\
        \hline
        \end{tabular}
    \end{center}
    \end{table}

%\begin{table}
%    \caption{Spatial scale of wavelet planes expressed in the {\nu} DET1 pixel (`pix'), physical (`mm'), and angular sizes ($''$).}
%    \label{tab:sizes}
%    \begin{tabular}[width=\textwidth]{l|r|r|r|r|r|r|r|r|r|r}
%    \hline
%        Scale & 
%        1 & 2 & 3 & 4 & 5 & 6 & 7 & 8\\
%        \hline
%        Size, pix &
%        2 & 4 & 8 & 16 & 32 & 64 & 128 & 256\\
%        %\hline
%        Size, mm &
%        0.24 & 0.48 & 0.96 & 1.92 & 3.84 & 7.68 & 15.36 & 30.72\\
%        %\hline
%        Size, $''$ &
%        4.92 & 9.84 & 19.68 & 39.36 & 78.72 & 157.44 & 314.88 & 629.76 \\
%    \end{tabular}
%\end{table}

The main goal of the proposed method is the detection and removal of the focused component. As a first step, we determine significant excess in each wavelet layer 3-8 and define the removal mask region. Second, we compare the signal in each pixel of the wavelet layer with the specific threshold level calculated for each pixel on each wavelet scale. The detailed description of this procedure is given in Appendix \ref{app:threshold}.

The choice of threshold is mainly determined by the goal of finding a significant signal without false positives. However, setting a sufficiently high filtering threshold on an image with extended features will result in the exclusion of only a peak of the extended object leaving broad wings not removed. This is especially important for the {\nu} focused signal because of the wide PSF wings. To overcome this, we apply two-step conditional threshold filtering on each wavelet plane $w_i(x, y)$.

First, we find the region with image pixels above the initial threshold level $t_{\rm max}$. Then we iteratively search for the neighbouring pixels around the detected region which exceed a different, lower threshold $t_{\rm min} < t_{\rm max}$ and expand region until no new pixels can be added. Note that too high $t_{\rm max}$ will not trigger detection of the structures, and too low $t_{\rm min}$ will select image regions not related to the structures of interest. After testing different combinations of thresholds, we found that configuration with $t_{\rm max}=3$ and $t_{\rm min}=2$, confidently detects focused X-ray signal in the {\nu} images on all wavelet scales $3-8$. Figure~\ref{fig:wav_layers} shows the result of conditional threshold filtering of the {\nu} reference image.

%{\nu} image (ObsID 30001019002, FPMB) on each scale.

\subsection{Removal of focused X-ray flux from the {\nu} images}
\label{filtering}
The removal of the focused X-ray component from the {\nu} detector images is achieved by masking out the area of the detector containing the significant X-ray excess.
% \footnote{An interesting alternative could be not a simple removal of the focused X-ray flux, but instead taking its morphology into account. To do this, the sum of wavelet scales 3-8 can serve as a model, describing the focused X-ray flux.}.

Applying conditional threshold filtering to each wavelet plane $w_i(x, y)$, we generate an image mask that contains zeroes at the position of the significant structures and ones in all other detector regions, thereby excluding the areas containing significantly detected structures.

In most cases, an image mask of a given scale is located within the mask area of a larger scale, as it is shown in Fig.~\ref{fig:wav_layers}. As scale increases, the more source-affected area is excluded and less detector is left for stray light analysis.

The method is designed to achieve a balance between the amount of clean detector area and maximum filtering out the focused component. To find the best solution for a given observation we suggest iterating over all possible combinations of wavelet masks optimizing the combination of two parameters: 1) fraction of clean detector area $S = N_{\rm cl}/N_{\rm tot}$, where $N_{\rm cl}$ and $N_{\rm tot}$ -- clean and total detector area, respectively; and 2)  statistical quality of the remaining detector area, expressed by the modified Cash-statistics \citep{2017Kaastra}.

Our hypothesis is that all detector pixels follow Poisson distribution with expected count rate $I_{\rm mean}$ constant across the detector. We should note that the assumption about a constant count rate across the {\nu} detectors is a strong one due presence of spatial gradient caused by aperture CXB, which constitutes the majority of the {\nu} stray light background beyond the Galactic plane. We performed detailed modeling of the detector background including flat instrumental component and aperture CXB flux and found $10\%$ systematic bias of Cash-statistic, which, however, does not affect the detection of spatial structures of interested scales. Thus, we chose to keep a simple assumption of flat detector background, which provides a robust solution for detection of spatial structures.

Prior to calculating Cash-statistic, we grouped pixels into bins containing at least 2 counts. The modified Cash-statistic per bin is calculated as:
\begin{equation}
\label{eq:cash}
    C = \frac{2}{N_{\rm bin}} \sum_{i} E_i - C_i + C_i \cdot (\log{C_i} - \log{E_i}) 
\end{equation}
where $C_i$ is the number of counts for each bin; $E_i$ is the expected value for each bin, calculated as the expected value for a single detector pixel $I_{\rm mean}$ multiplied by the number of pixels in the bin; $N_{\rm bin}$ is the total number of bins. In this configuration, the remaining focused X-ray flux on the detector will produce worse Cash-statistic adding non-statistical systematic noise to the data.

Our goal now is to choose the combination that provides the optimal ratio between maximally removed focused X-ray flux and the area of the clean detector suitable for stray light analysis. For a given observation, we calculate $S$ and $C$ for all possible combinations of wavelet plane masks. For example, for 6 wavelet layers, that is, for 6 different masks, the total number of possible combinations would be $2^6 = 64$ combinations. The final mask for each combination is obtained by applying the logical \texttt{OR} operation between the selected image masks, which exclude structures of selected spatial scale.

As an initial guess, we apply all generated masks, which provide the smallest fraction of clean detector area $S_0$ but maximally exclude the contribution of detected sources, resulting in some baseline Cash-statistic value $C_0$. Then we iterate over all other mask combinations and search for the optimal one with the biggest fraction of clean detector $S$ while the Cash-statistic $C$ is below $C_0 + 0.05$.

%\rk{\sout{We first choose the combination with the smallest fraction of clean detector $S_0$ (i.e. maximally removed focused X-ray flux) and whatever Cash-statistics $C_0$. Then we consider all other mask combinations that satisfy the following condition for $C$ and $S$:}}
% \begin{equation}
%     \begin{cases}
%     C < C_0 + 0.05\\
%     S > S_0
%     \end{cases}
%     \label{eq:rules}
% \end{equation}

In other words, we try to increase the area of the clean detector, allowing some little worsening of the Cash-statistic. The threshold value of 0.05 is chosen to roughly approximate the $1\sigma$ threshold for this statistic\citep{2017Kaastra}. As a result, we choose the optimal combination of wavelet masks that provide the largest area of clean detector $S_{\rm opt}$ along with acceptable Cash-statistic $C_{\rm opt}$. In case no mask combinations satisfy given conditions, we accept the initial assumption $S_0$ and $C_0$ as the optimal one ($S_{\rm opt}$, $C_{\rm opt}$).

\begin{figure}
    \centering
    \includegraphics[width=\linewidth]{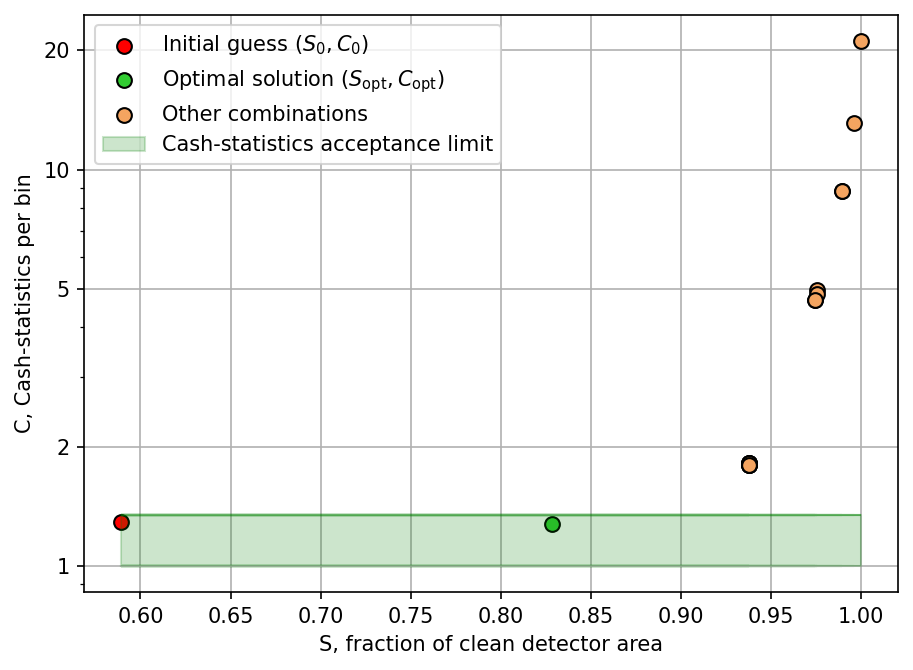}
    \caption{Illustration to the iterative algorithm to choose the wavelet plane mask combinations for reference {\nu} observation with a bright source in the FOV (see Fig.~\ref{fig:observation}). Points illustrate the fraction of clean detector area $S$ and Cash-statistic $C$ for different combinations of masks. The green point corresponds to the combination chosen by the algorithm as the optimal solution. The red point shows the initial guess. Yellow points represent all remaining mask combinations. Note that due to nesting of the wavelet masks, the yellow points can contain many mask combinations.}
    \label{fig:param}
\end{figure}

Figure~\ref{fig:param} shows an illustration of the detector area optimization for the reference {\nu} observation (see Figs.~\ref{fig:observation}, \ref{fig:wav_layers}). Obviously, the combination with maximally removed focused X-ray flux provides a minimal clean detector area. As we apply the mask combinations with smaller masked area, the raising fraction of focused X-ray flux on the detector rapidly worsens Cash-statistic until reaching the maximum, when no mask is applied to the data. However, thanks to the optimization procedure, we select the best solution with a clean detector area fraction of ${\approx}0.83$ instead of ${\approx}0.58$ (the initial assumption), increasing the detector area suitable for stray light analysis (see also Sect.~\ref{sec:archive} for demonstration on a large data set).

\begin{figure}
    \centering
    \includegraphics[width=\linewidth]{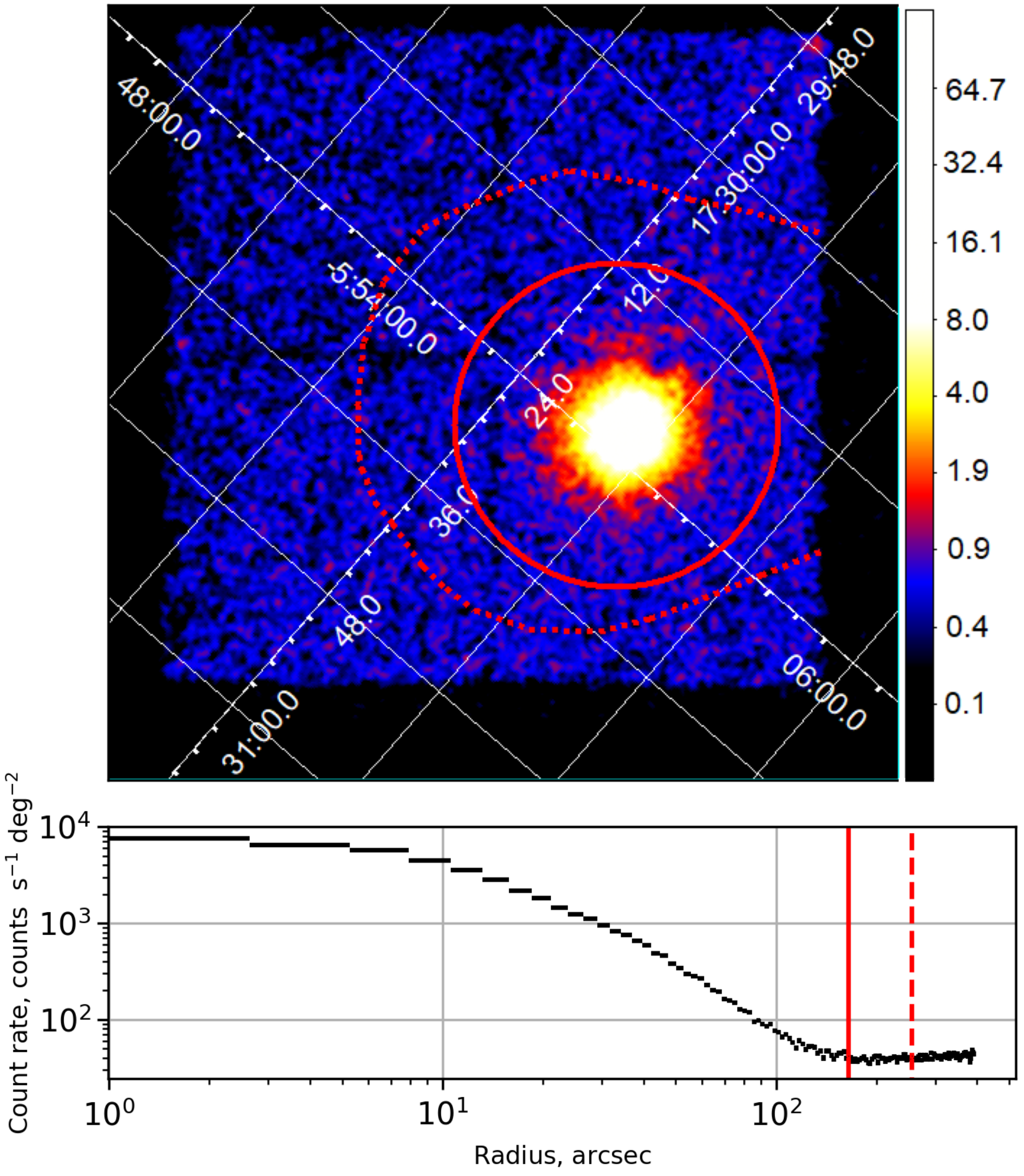}
    \caption{{\it Top}: The {\nu} reference image (Fig.~\ref{fig:observation}) in SKY coordinate system. Red dashed and solid regions represent the initial and optimal wavelet mask solutions, respectively. {\it Bottom}: Radial profile of the source centered at the peak position. Vertical lines show the approximate source exclusion radii obtained for initial (dashed) and optimal (solid) mask solutions, similar to the image above.}
    \label{fig:example1}
\end{figure}

Visual inspection of the observation and generated masks in the SKY coordinate system (reconstructed sky image) further reinforces the difference between the initial guess and optimal mask. Fig. \ref{fig:example1} shows that, both the initial and optimal wavelet mask solutions provide a sufficient removal of the focused X-ray signal. Since the on-axis image of a point-like source in SKY coordinate system is characterized by a single and symmetric peak, we extracted an unbiased radial profile of a source shown in bottom of Fig. \ref{fig:example1}. As seen from the profile, the background starts to dominate at angular offsets ${\gtrsim}200''$ from the source position and wavelet-based method provides guaranteed (initial) and well-determined (optimal) source exclusion radius.

\section{Application examples}

In this section we present the efficiency of the method applied to a selected {\nu} observations with known image artifacts and to a large number of observations from the {\nu} archive.

\subsection{{\nu} observations with artifacts}

\begin{figure}
    \centering
    \includegraphics[width=\linewidth]{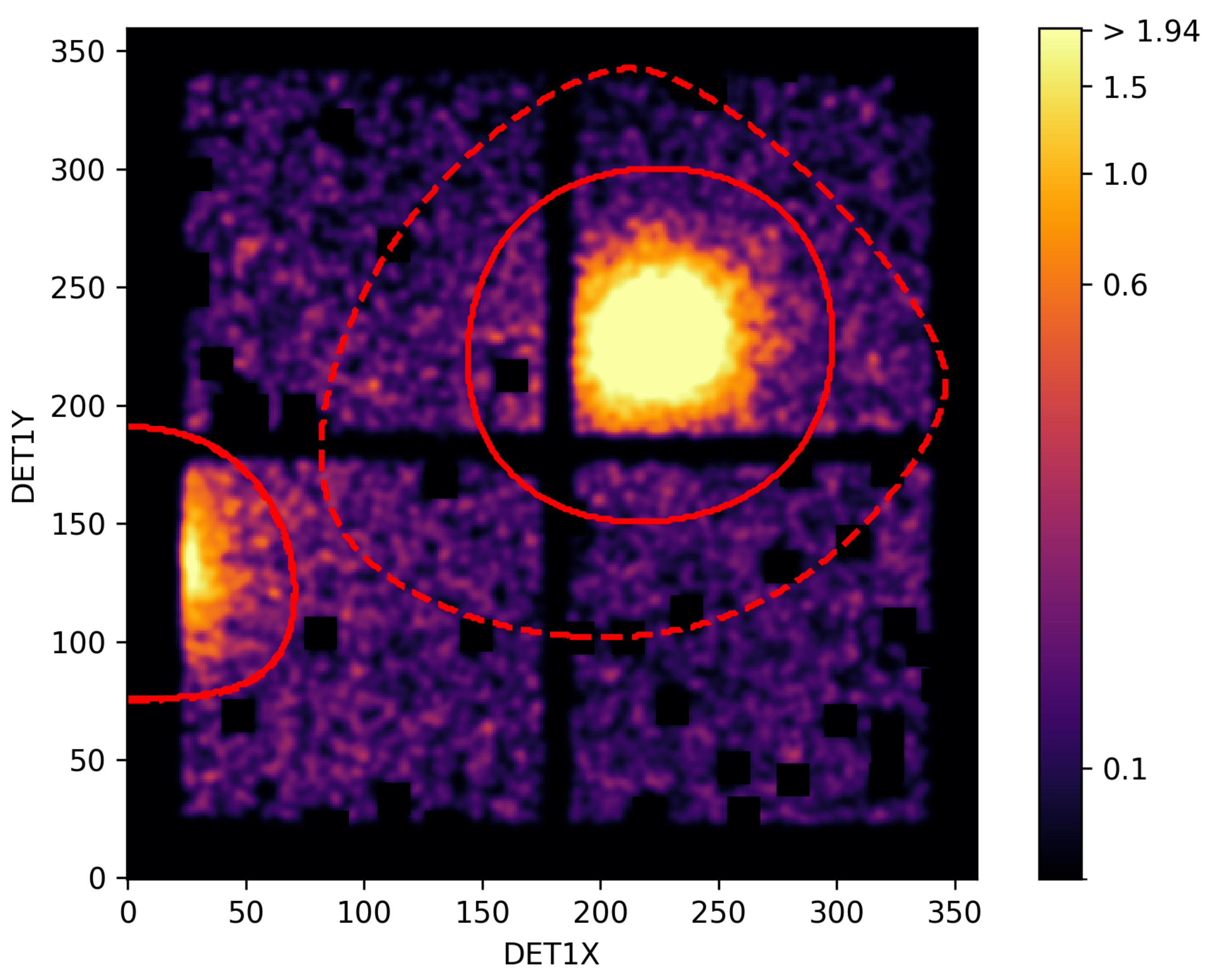}
    \caption{The {\nu} 3--20 KeV image of a bright Gamma-ray binary PSR~B1259--63 (ObsID 30002017010, FPMA, DET1 coordinate system). Red dashed and solid contours represent initial and optimal wavelet mask solutions, respectively.}
    \label{fig:example2}
\end{figure}

The first example shown in Fig.~\ref{fig:example2} is the on-axis $30$~ks observation of a bright Gamma-ray binary PSR~B1259--63 \citep{2015MNRAS.454.1358C} contaminated with a ${\sim}5$ times weaker ghost-rays from a large off-axis source 2RXP~J130159.6--63580 \citep{2015ApJ...809..140K}. The method successfully removes flux from PSR~B1259--63 in the center of the image and successfully determines the contaminated region at the edge of the detector.

\begin{figure}
    \centering
    \includegraphics[width=\linewidth]{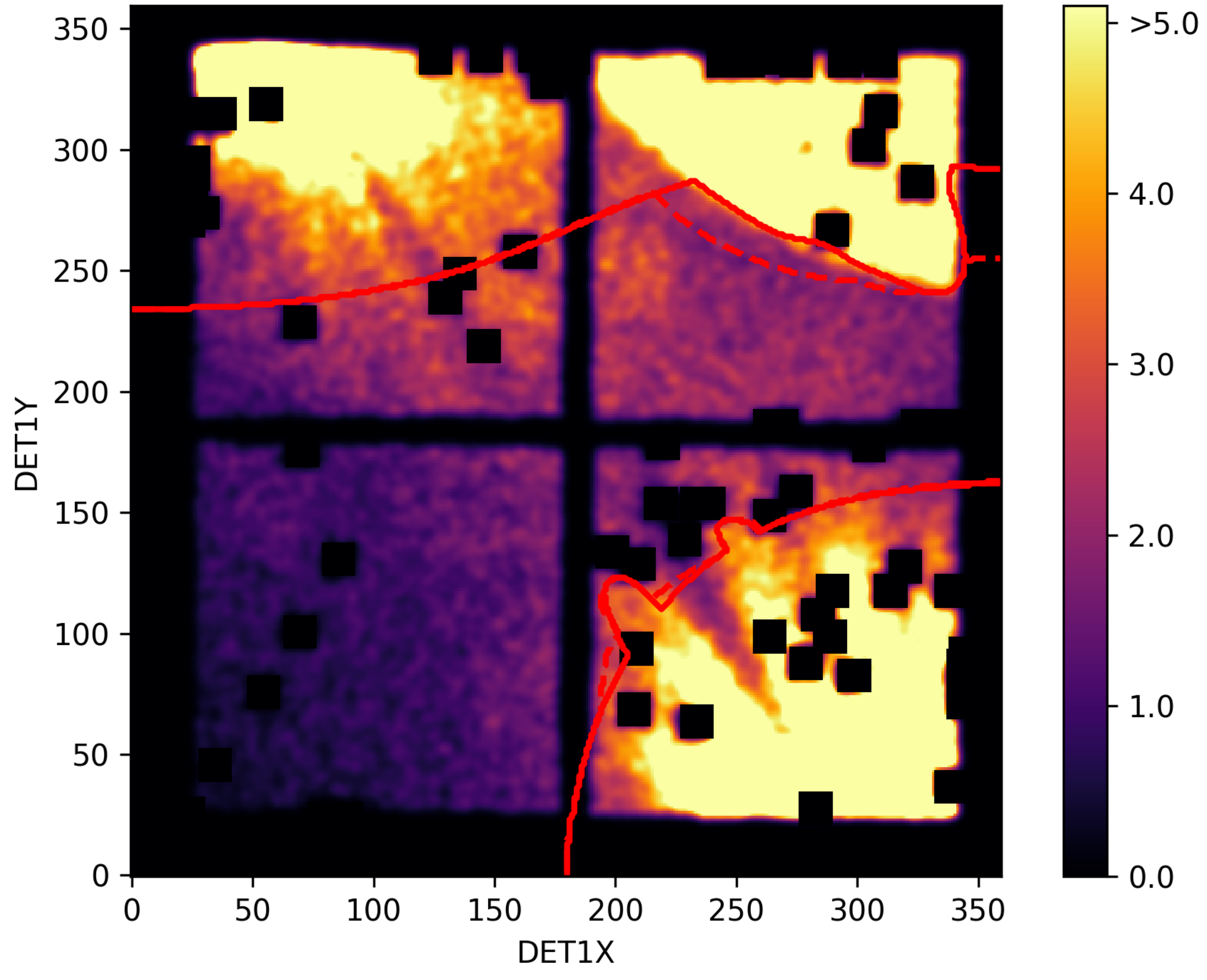}
    \caption{The {\nu} 3--20 KeV image of the observation with bright ghost-ray and stray-light contamination (ObsID 90201020002, FPMB, DET1 coordinate system). Red dashed and solid contours represent wavelet initial and optimal wavelet mask solutions.}
    \label{fig:example3}
\end{figure}

The second example (Fig. \ref{fig:example3}) is a complex 90~ks observation taken on 2016 September 19 having a number of known {\nu} image artifacts used for demonstration in dedicated article \citep{2017Madsen}. The FPMB detector image contains bright focused flux from two far off-axis sources, contaminated by their ghost rays (top-left and bottom-right corners), and strong stray-light from a nearby bright source (top-right corner). Despite the strong contamination, the method provides relatively well-determined regions of a clean detector area. One can notice at least two main features on this figure. First, the mask optimization algorithm allocated additional clean detector area below the top-right stray light. Second, the method shows incomplete determination of the top-left structure caused by ghost rays. This observation reveals the main limitations of wavelet-based methods, as they are unable to detect highly asymmetric structures or structures with sharp edges.

\subsection{Method properties for a large data set}
\label{sec:archive}

To investigate efficiency and properties of the method on a large data set, we chose 6573 {\nu} observations with exposure more than 1000~s spanning 10 years of operations from 2012 July 1 to 2022 March 13. 

%In this section, we present the statistical properties of a large set of almost 10 years of observations of {\nu} to showcase the possibilities of the method. Here we used the observation products from the {\nu} Master Catalogue archive from 2012 July 1 to 2022 March 13.

\begin{figure}
    \centering
    \includegraphics[width=\linewidth]{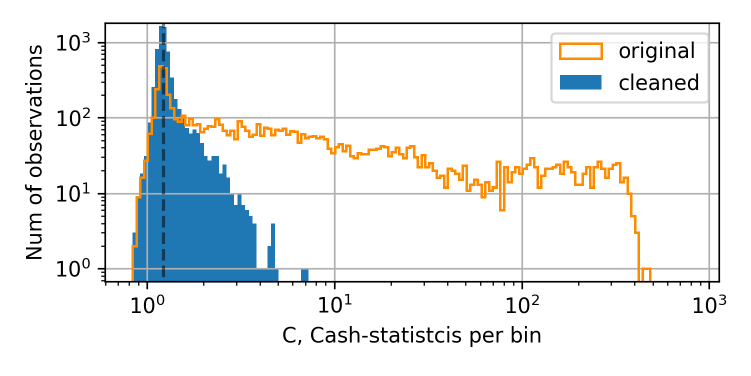}
    \caption{Histogram of modified Cash-statistics per bin $C$ for 6573 {\nu} observations. The orange and solid blue histograms show the distributions of $C$ values for original observations ``as is'' and detector images cleaned with the presented method, respectively. Vertical dashed line represents median value $C=1.2$ for cleaned data.}
    \label{fig:cash_stat}
\end{figure}

Figure~\ref{fig:cash_stat} shows histogram of modified Cash-statistics $C$ (Equation~\ref{eq:cash}) for {\nu} observations with a wide range of target source brightness processed with our method. The maximum value of $C$ is 7.0 for cleaned observation with ObsID 10110004002 contaminated with stray light from a bright source covering ${>}50\%$ of FPMA detector area. The relative fraction of observations with $C<2.0$ (which roughly corresponds to a good statistical quality for stray-light background studies) is 35\% and 94\% for, respectively, original (uncleaned) and source-free (cleaned) observations. 

%94\% of observations after the processing have Cash-statistics per bin below 2.0 compared to only 35\% of observations before processing.

%The main metric of the data representing the goodness of the masking procedure is the modified Cash-statistics per bin (Equation~\ref{eq:cash}). As shown in Figure \ref{fig:cash_stat} all of the observations with bright sources are processed with the algorithm. 94\% of observations after the processing have Cash-statistics per bin below 2.0 compared to only 35\% of observations before processing.

\begin{figure}
    \centering
    \includegraphics[width=\linewidth]{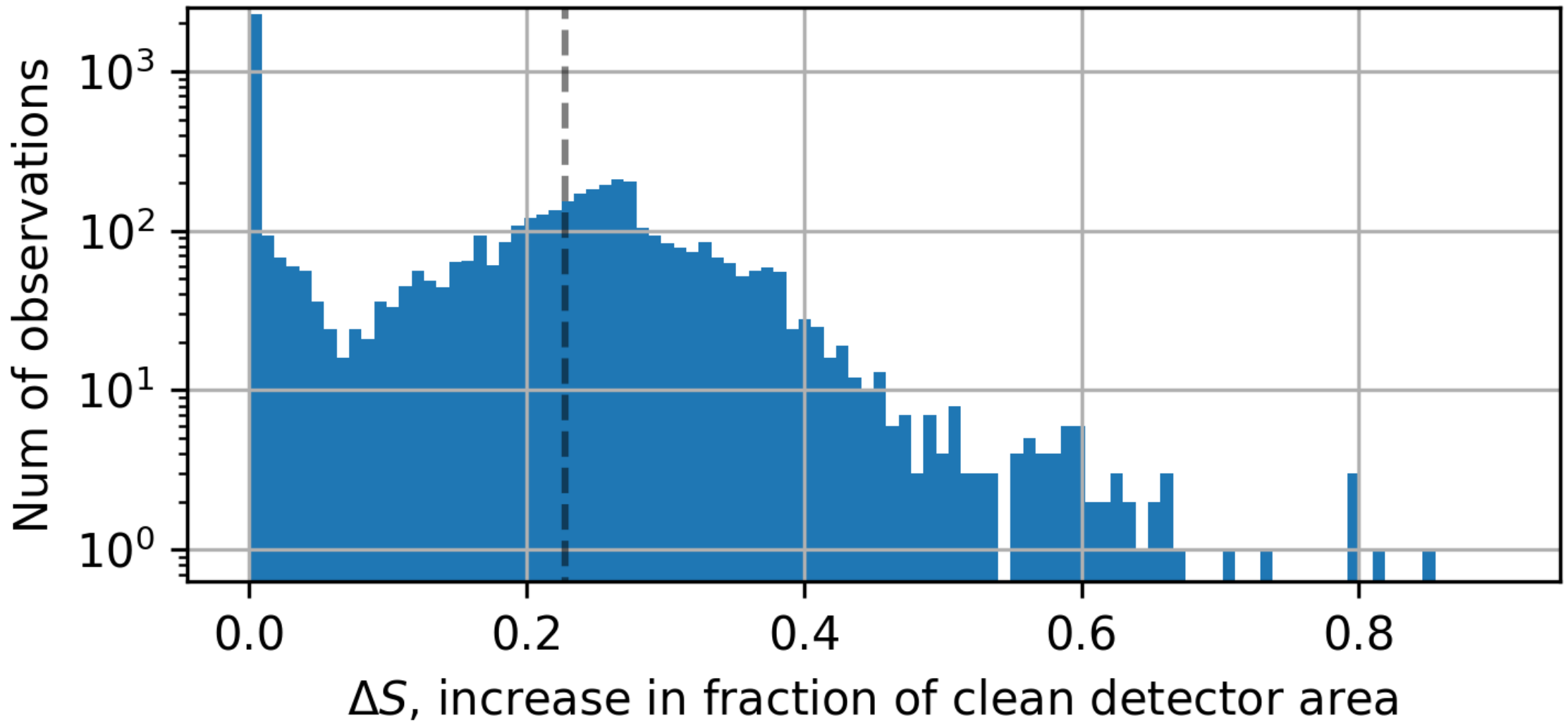}
    \caption{Histogram of clean detector area increase $\Delta S=S_{\rm 0}-S_{\rm opt}$ as a result of wavelet mask optimization for 6573 {\nu} observations. Vertical dashed line shows median value  $\Delta S=0.23$, excluding bin with  $\Delta S=0$.}
    \label{fig:fraction}
\end{figure}

Then, we evaluated the efficiency of detector area optimization, i.e. the choice of optimal wavelet mask combination ($S_{\rm opt}$, $C_{\rm opt}$) with respect to the initial setup ($S_{\rm 0}$, $C_{\rm 0}$). We defined efficiency in terms of a difference of detector area $\Delta S=S_{\rm 0}-S_{\rm opt}$. The histogram of $\Delta S$ is shown in Figure~\ref{fig:fraction}. We calculated the median value of $\Delta S$, excluding bin with $\Delta S=0$, because it  corresponds to blank-sky detector images without X-ray sources. The resulting median value is 0.23, which means that optimization procedure adds in average more than ${\sim}20\%$ of clean detector area for stray-light background studies.

%The metric representing the efficiency of our optimal mask choosing algorithm is the fraction of the clean detector. In Figure \ref{fig:fraction} we present the histogram with the difference of said metric between the initial mask and optimal mask combination for the same set of observations. The resulting mean difference in the fraction of clean detector area is 0.13, giving a 23\% increase compared to the initial guess masks, while the difference in the Cash-statistics cannot exceed 5\% due to the restraints put on the algorithm. 

\section{Summary}
The {\nu} side aperture (or stray light) is a known source of rich astrophysical information. However in order to take advantage of it, one need to characterize and remove focused X-ray flux, often, for a large number of {\nu} observations.

In order to support many {\nu} stray light related studies, we present fully automatic method for cleaning detector images from any kind of focused X-ray flux, while maximizing the detector area suitable for analysis. The main idea of the method is `\'a trous' wavelet linear image decomposition, capable of detecting structures of any spatial scale and shape. Note that the presented method has a general application for any astrophysical X-ray images. 

For {\nu} data, a number of settings can be applied to customize the method for a specific use. First, we introduce conditional thresholds to detect structures of different spatial scales. Second, we implement the iterative procedure for optimizing the amount of cleaned detector area, which additionally allocates ${\sim}20\%$ of detector area in average. As a result, the method provides  detector image area with the highest possible statistical quality, suitable for the {\nu} stray light background studies.

%The last It opens the possibility of balancing the preservation of maximum cleaned detector area and maximum source signal exclusion, providing the result with the best statistical properties for further stray light analysis.

%We expanded the general approach with a set of additional adjustments to provide a more {\nu} specific solution to the problem. Firstly, we perform a two-step conditional threshold to include the wide spread of the sources. Secondly, we realised a simplistic iterative method for optimized detector area filtering. It opens the possibility of balancing the preservation of maximum cleaned detector area and maximum source signal exclusion, providing the result with the best statistical properties for further stray light analysis.

%\footnote{\url{https://github.com/NuSTAR/nuwavdet}}$^{,}$

We developed open-source Python \texttt{nuwavdet} package (See Code and Data Availability), which implements the presented method and provides separation of the focused X-ray flux and stray light background for the {\nu} observations. The package contains subroutines to generate detector image regions with clean background, which can be used for processing within the Python programming environment, and optionally, to save them as Flexible Image Transport System (FITS) files. In particular, the resulting detector regions can be stored as a list of user-defined bad pixels, suitable for processing in the NuSTARDAS for conventional X-ray analysis.

\section*{Code and Data Availability}
The code developed for this work is publicly available in \url{http://heagit.cosmos.ru/nustar/nuwavdet}. This work is based on publicly available data of the {\nu} telescope. The High Energy Astrophysics Science Archive Research Center (HEASARC, \url{https://heasarc.gsfc.nasa.gov}) provides the scientific archive of the \nu\ data to the community. 

\section*{Acknowledgements}

This work was supported by the Russian Science Foundation within scientific project no. 22-12-00271.

\subsection*{Disclosures}
The authors have no relevant financial interests and no other potential conflicts of interest to disclose.

%%%%% References %%%%%

\bibliographystyle{mnras}
\bibliography{refs}

% %%%%%%%%%%%%%%%%% APPENDICES %%%%%%%%%%%%%%%%%%%%%

\appendix
\section{{\nu} image preprocessing}
\label{app:prep}
Since the wavelet-based method presented in this article uses convolutional filters of various spatial scales, known {\nu} instrumental image artifacts  can introduce systematic noise into the wavelet space, making the method unstable. Among these are bad-flagged detector pixels and gaps between detector chips, which act in the processing as pixels with zero counts. The general approach to overcome this problem is to fill out the missing information with some expected counts.

The list of {\nu} bad pixels contains so-called stable bad pixels known at the on-ground calibration stage and pixels disabled by onboard processing. The latter are provided for each {\nu} observation with the corresponding time interval when the pixel was disabled. If a pixel was disabled for 10\% or less of the total observation time, we consider that pixel to be normal and populate it with the simulated Poisson counts as if it was working all the time. Otherwise, the pixel is considered as bad-flagged, and treated as follows.

We fill out bad pixels and detector gaps by the simulated Poisson counts with the expected value of a local mean, calculated as the mean of pixel values within a box of $15\times15$ pixels around a given position. In case the fraction of masked pixels within that box is more than 30\%, the wider box is taken. This procedure is repeated until all masked pixels are filled out.

As known from the analysis of the {\nu} data, the internal background count rate of each detector chip can be different by $5-10\%$ with respect to the mean level \citep{2021Krivonos}. Following the previously developed approach\citep{2021Krivonos}, we calculated relative chip normalizations listed in  Table \ref{tab:coeffs} using the {\nu} Earth occultation  observations carried out in 2013-2020.

To avoid boundary effects we perform image reflection at the image boundaries. To do this, the {\nu} detector image is copied into a 3 times bigger array where the central part represents the original image and all the surrounding parts contain reflected images.

\begin{table}
    \centering
    \begin{tabular}{c|c|c|c|c}
        \hline
        Focal plane & DET0 & DET1 & DET2 & DET3\\
        \hline
         FPMA & 1.047 & 1.166 & 0.979 & 0.861\\
         FPMB & 0.978 & 1.028 & 1.005 & 0.996
    \end{tabular}
    \caption{Table of normalizing coefficients for different detector chips}
    \label{tab:coeffs}
\end{table}

% %%%%%%%%%%%%%%%%%%%%%%%%%%%%%%%%%%%%%%%%%%%%%%%%%%

\section{Wavelet decomposition}
\subsection{General idea}
\label{app:wavelet}

\begin{figure}
    \centering
    \includegraphics[width=\linewidth]{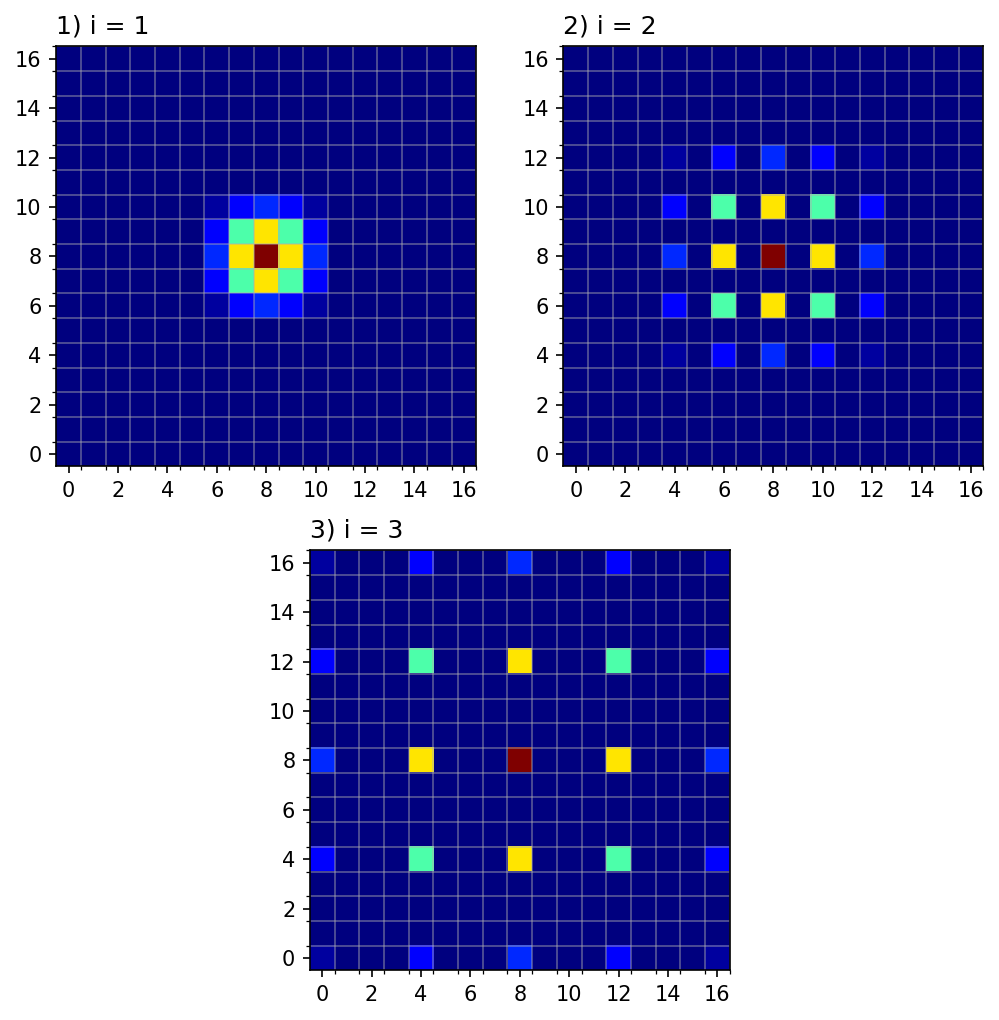}
    \caption{\'{A} trous kernel of different scales i: 1) i = 1, 2) i = 2, 3) i = 3.}
    \label{fig:atrous}
\end{figure}

The core idea of our algorithm is wavelet transformation. It is used to find spatial features of a given scale in the image. Conceptually wavelet transform is an operation of calculating the convolution of the image data I(x, y) with a series of wavelet kernels $\psi(\frac{x}{2^i}, \frac{y}{2^i})$ of different scales $i$. In discrete analysis the resulting image for a scale $i$ (wavelet layer of scale $i$, also referred hereafter as ``wavelet plane'') is calculated as:
\begin{equation}
        w_i (x, y) = (I\otimes \psi)(x, y) = \sum_{x', y'}{I(x-x', y-y') \cdot \psi(\frac{x'}{2^i}, \frac{y'}{2^i})}
\end{equation}
where $x, y$ are related to the pixel position on the image and $x', y'$ are pixel positions on the wavelet kernel matrix around its centre.

The resulting wavelet layers of scale $i$ provide information about the structures of a given spatial size $2^i$ in the image. As convolution is a linear operation we do not calculate the wavelet kernel itself. Instead, we perform two separate convolutions of the image with low pass filters of different scales and subtract the convolution with a bigger scale from the convolution with a lower scale.

Our algorithm uses the \'{a} trous low pass filter \citep{Holschneider}. On the scale 1 it is a $5 \times 5$ matrix with constant coefficients (Equation \ref{eq:atrous}) roughly approximating the Gaussian of width 2. The \'{a} trous low pass filter of scale {\it i} is calculated by inserting $2^{i-1}-1$ zeros between all elements of the original kernel, thus very roughly approximating the Gaussian with a width of $2^i$. The example of \'{a} trous kernels of different levels are shown in Figure \ref{fig:atrous}.
\begin{equation}
    \label{eq:atrous}
    h_1 (x, y) = \frac{1}{256}
    \begin{pmatrix}
    	1 & 4 & 6 & 4 & 1\\
    	4 & 16 & 24 & 16 & 4\\
    	6 & 24 & 36 &24 & 6\\
    	4 & 16 & 24 & 16 & 4\\
    	1 & 4 & 6 & 4 & 1\\
    \end{pmatrix}
\end{equation}

Then, we introduce so-called wavelet layer at scale $i$ described by the following equation:

\begin{equation}
    w_i (x, y) = (I \otimes h_{i-1})(x,y) - (I\otimes h_i)(x, y)
    \label{eq:wavelet}
\end{equation}

Here the $h_0(x,y)$ is a $\delta$-function (1 in (0,0) position and 0 otherwise) which is effectively leaving the image as is.

One of the disadvantages of the basic wavelet transformation is that signals of smaller spatial scales (high frequency, e.g. point-like sources) can interfere with the wavelet layer of the larger spatial scales (low frequency, e.g. extended emission). To overcome this we iteratively subtract small scale features from the image before calculating the wavelet layer of a larger scale. Thus, the method consists of the following steps.

In the first step, we perform wavelet transformation according to Equation \ref{eq:wavelet} for scale 1 on the image $I_0(x,y) = I(x,y)$. This gives us the wavelet layer $w_1(x,y)$.
\begin{equation}
    w_1 (x, y) = (I_0 \otimes h_{0})(x, y) - (I_0 \otimes h_1)(x, y)
\end{equation}

Then this wavelet layer is subtracted from the data, creating new image $I_1(x, y)$:
\begin{equation}
    I_1(x, y) = I_0(x, y) - w_1(x,y)
\end{equation}

In the second step, we repeat the same procedure with image $I_1(x, y)$ using the wavelet at scale 2 and obtain the wavelet layer  $w_2(x,y)$. This process is repeated until we reach the maximum number $n$ of desired wavelet layers.

In the final step, $n$ we perform the convolution of the image $I_{n-1}(x, y)$ with the low pass filter of the scale $n-1$ and obtain final layer $c(x,y)$:
\begin{equation}
    c(x, y) = (I_{n-1} \otimes h_{n-1})(x, y)    
\end{equation}

As a result, we obtain a set of wavelet layers $w_i(x, y)$ at different spatial scales and a final layer $c(x, y)$ with the remaining large-scale features above scale $n$. 

The main concept of the whole method is that the obtained wavelet planes constitute a linear decomposition of the original image:
\begin{equation}
    I(x, y) = \sum^{n-1}_{i=1} w_i(x, y) + c(x, y)
\end{equation}

Thus, we can consider $w_i$ as images containing ``flux'' on scale $i$. To reveal structures of a certain scale in an image, one simply needs to take the sum of the corresponding wavelet layers where the structure of interest is most noticeable. For example, point sources are mainly visible at small scales $i$, while extended structures are predominantly seen at large values of $i$.

\subsection{Conditional threshold filtering}
\label{app:threshold}
In the previous section, we described a general approach to perform linear decomposition of the 2D data, and in particular, revealing X-ray focused component on the {\nu} detector images. In order to remove the observed structures on each wavelet layer, one needs to determine the level above which the signal is statistically significant. Here we generally treat the data as having the Gaussian noise distribution after modifying uncertainty following Poisson low-count statistics.

To estimate the threshold for significant detection let us first assume the noise with Gaussian distribution and constant $rms=1$ across the image. The distribution of pixel values after convolution with kernel $\psi(\frac{x}{2^i}, \frac{y}{2^i})$ is then also Gaussian with the root mean square value scaled from 1 by the factor of $\sigma_{i}$, which can be calculated analytically as:

\begin{equation}
    \sigma_i = \sqrt{\sum_{x,y} \psi^2(\frac{x}{2^i}, \frac{y}{2^i})}
\end{equation}

In our case, in order to approximate the residual of two \'{a} trous low pass filters we utilize the difference of two discrete Gaussian curves $g_i(x, y)$ of the corresponding width $2^i$. Thus, the value of $\sigma_i$ for our algorithm is:
\begin{equation}
    \label{eq:sigma}
    \sigma_i = \sqrt{\sum_{x,y} (g_{i-1}(x, y) - g_{i}(x, y))^2}
\end{equation}

\begin{figure}
    \centering
    \includegraphics[width=\linewidth]{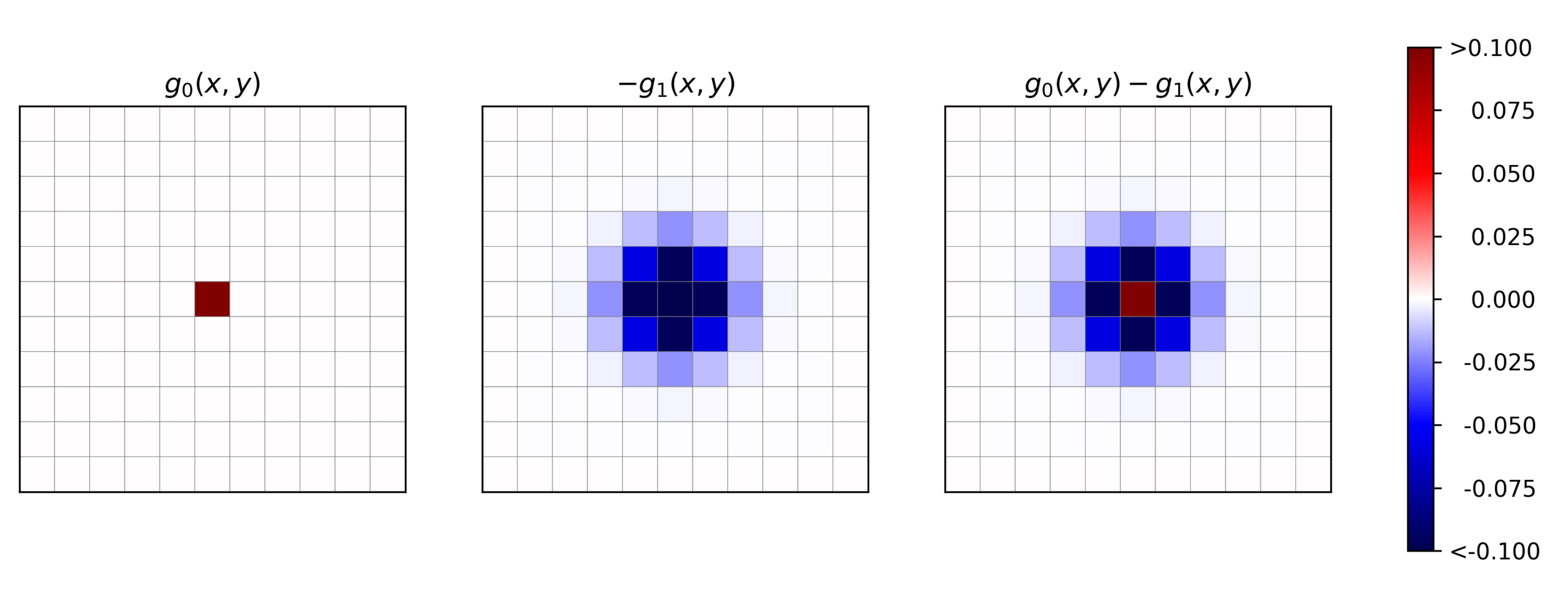}
    \caption{An illustration to the Equation \ref{eq:sigma} for $\rm i = 1$. The left image illustrates the discrete Gaussian of size 0, central image shows the negative Gaussian of size 1, the right image shows the result of element-wise subtraction of the given Gaussians.}
    \label{fig:gauss_example}
\end{figure}

For example, for $\sigma_1$ we calculate the element-wise difference of two arrays corresponding to the Gaussian with zero size (which is treated as a kernel, not modifying the original data) and a Gaussian with the scale of 1 as shown in Fig. \ref{fig:gauss_example}.

\begin{table}
    \caption{$\sigma$ values  for \'{a} trous wavelet transform calculated with Equation \ref{eq:sigma}.}
    \label{tab:atrous}
    \begin{tabular}{ cllllll}
        \hline
        i & 1 & 2 & 3 & 4 & 5 & n\\
        \hline
        $\sigma_i$ & 0.8725 & 0.1893 & 0.0946 & 0.0473 & 0.0237 & $\sigma_{n-1}/2$
    \end{tabular}
\end{table}

If the noise in the original data has $rms = k$ we can calculate the root mean square of the image after convolution with the kernel of scale $i$ as $k\cdot \sigma_i$. We are generally using this value as the 1$\sigma$ threshold level, assuming that root mean square across the image is constant.

Then we assume that data is distributed as a Poisson signal. Following the Gehrels' work\citep{1986ApJ...303..336G} we approximate  the Poisson $1\sigma$ error of data as:
\begin{equation}
    k_i(x, y) = 1 + \sqrt{bkg(x, y) + 0.75}
\end{equation}
where $bkg(x, y)$ is the approximation of the background signal that is estimated as the convolution of data with an \'{a} trous low pass filter of a Gaussian with width $2^{i}$.

Thus, the t-level confidence detection after additionally performing the wavelet transformation is described by:
\begin{equation}
    w_i(x, y) > t\cdot k_i(x, y) \cdot \sigma_i
\end{equation}

% \disclosures
\end{document}